\documentclass[prl,showpacs,floatfix,twocolumn]{revtex4}
\usepackage{graphicx}

\makeatletter
\newbox\slashbox \setbox\slashbox=\hbox{\large$/$}
\def\pslash#1{\setbox\@tempboxa=\hbox{$#1$}
  \@tempdima=0.5\wd\slashbox \advance\@tempdima 0.5\wd\@tempboxa
  \copy\slashbox \kern-\@tempdima \box\@tempboxa}
\def\slash{\protect\pslash}
\makeatother
\newcommand{\be}{\begin{eqnarray}}
\newcommand{\ee}{\end{eqnarray}}
\def\beq{\begin{equation}}
\def\eeq{\end{equation}}
\def\SB{S$\chi$SB}
\def\fm{\,\mathrm{fm}}

\begin{document}
\title{The QCD vacuum as a disordered medium:\\
A simplified model for the QCD Dirac operator}
\author{Antonio M. Garc\'{\i}a-Garc\'{\i}a}
\affiliation{Laboratoire de Physique Th\'eorique et Mod\`eles 
Statistiques, B\^at. 100, \\ Universit\'e de Paris-Sud, 
91405 Orsay Cedex, France}
\author{James C. Osborn}
\affiliation{Physics Department, University of Utah,
Salt Lake City, UT 84112, USA}
\begin{abstract}
We model the QCD Dirac operator as a power-law random banded matrix 
(RBM) with the appropriate chiral symmetry.
Our motivation is the form
 of the Dirac operator in a basis of instantonic zero modes with
 a corresponding gauge background of instantons.
We compare the spectral correlations of this model to those of an instanton
liquid model (ILM) and find agreement well beyond the Thouless energy.
In the bulk of the spectrum the (dimensionless) Thouless energy of the
RBM scales with
the square root of system size in agreement with the ILM and
chiral perturbation theory.
Near the origin the scaling of the (dimensionless) Thouless energy in the
RBM remains the
same as in the bulk which agrees with chiral perturbation theory but not
with the ILM.
Finally we discuss how this RBM should be modified in order to 
 describe the spectral correlations of the QCD Dirac operator at 
 the finite temperature chiral restoration transition.
\end{abstract}
\pacs{72.15.Rn, 71.30.+h, 05.45.Df, 05.40.-a} 
\maketitle

In recent years  
the relation between spontaneous chiral symmetry breaking
(\SB) in QCD and the phenomenon of conductivity in a disordered medium has
 been investigated in the literature \cite{diakonov1,VO,zahed}.
In the latter, conductivity is produced by electrons that although
initially bound to impurities may get delocalized by orbital overlapping. 
In QCD the quark zero modes 
initially bound to an instanton get delocalized due to  
the strong overlap with other would-be
 zero modes and consequently chiral symmetry is broken.
In the case of atoms the overlap 
is effective only among nearest neighbors due to the 
exponential decay of the electron wavefunction.
 However, in the QCD vacuum, 
 the decay is power-law and long range hopping is possible. 
Thus even if one assumes a perfect 
 trade between impurities and instantons and between
 electrons and quarks, the associated 
 Anderson model should posses long range hopping.   
 Such models have already been discussed in the 
literature \cite{ono,levitov1}.
The main conclusion of these works was that power-law 
 hopping may induce a metal insulator transition
  even in one dimensional systems if the exponent of the 
 hopping decay matched the dimension of the space. 
 Similar findings were also reported in a related model, 
a random banded matrix 
with a power-law decay \cite{par,prbm}.

Random matrix techniques have already been used in the context of QCD.
In the infrared limit the 
 eigenvalue correlations of the QCD Dirac operator
 do not depend on the dynamical details of the QCD Lagrangian
  but only on the global symmetries 
  of the QCD partition function \cite{SV}. 
Thus random matrices with the correct chiral symmetry
 of QCD (termed chiral random matrices) \cite{V} accurately describe 
  the spectral properties of the QCD Dirac operator up to some scale 
known as the Thouless energy. 
For larger energy differences dynamic features become important and the
standard random matrix model ceases to be applicable.

In this letter we study an improved matrix model for QCD which reproduces
features of the eigenvalue spectrum beyond the Thouless energy.
We want to incorporate the phenomenological
 fact that the matrix elements of the QCD Dirac  
 operator in a basis of zero modes  
 decay as a power of the instanton--anti-instanton distance
 into the random matrix approximation.
We thus propose  
 a chiral RBM 
 as an effective model of the QCD
 Dirac operator.  
We first give a brief introduction on the role of instantons in the QCD vacuum.
Next we introduce the chiral RBM and describe its main properties
at both the origin and bulk of the spectrum.
In the latter case an analytical treatment is possible. 
Then we compare the spectral correlations of this model 
with  the results of an instanton liquid simulation.
Finally we discuss how the RBM can be modified 
to describe the spectral correlations 
of the QCD Dirac operator at the finite temperature chiral restoration
transition.

\section{\SB{} from Instantons}

The discovery of instantons has had a large impact on our understanding
of non-perturbative aspects of QCD \cite{polyakov}
 (for a modern review see \cite{SS97}).
Instantons are classical solutions
 of the Yang-Mills equations of motion in Euclidean space.
An important property is 
that the Euclidean QCD Dirac operator has an
 exact zero eigenvalue in the field of an instanton. 
The spectral properties of the low lying modes of the Dirac
 operator are thus controlled by these 
 non-perturbative configurations.

Unfortunately the construction of a 
 consistent QCD vacuum based on instantons 
faces serious technical difficulties.
Exact analytical multi-instanton configurations  
 are hard to obtain since the Yang-Mills equations of motion for QCD
 are nonlinear and therefore a superposition 
of single instanton contributions is not itself a solution.
Additionally quantum corrections may spoil 
the semiclassical picture implicitly assumed
 of a QCD vacuum composed of instantons well 
 separated and weakly interacting.   
  These problems  
have been overcome either by 
invoking variational principles \cite{diakonov} or by phenomenologically 
fixing certain parameters of the instanton ensemble.
The latter case, usually referred to 
 as the instanton liquid model (ILM) \cite{shuriak}, yields
 accurate estimates of vacuum condensates and hadronic
 correlation functions \cite{SSV} just by setting
 the density of
 instantons to be $N/V \approx 1 \fm^{-4}$ and the mean size 
 ${\bar \rho} \approx 1/3 \fm$.
Lattice simulations have also supported the picture
 of a QCD vacuum dominated by instantons \cite{lat}.

An advantage of these models is that they
satisfactorily explain how chiral symmetry 
 is spontaneously broken in the QCD vacuum \cite{diakonov}. 
The chiral condensate
 (the order parameter signaling \SB)  
  $\langle {\bar \psi} \psi \rangle$
 is related to the spectral 
density $\rho(\epsilon)$ of the QCD Dirac operator around zero
 through the Banks-Casher
 relation \cite{bank},
\be
\langle {\bar \psi} \psi \rangle =
 - \lim_{\epsilon \to 0} \lim_{V \to \infty} \frac{\pi \rho(\epsilon)}{V} ~,
\ee 
where $V$ is the space-time volume.
The ILM provides a phenomenological model for these low energy modes of the
QCD Dirac operator.
In a basis of $N/2$ instantonic zero modes $\psi_{I}({\vec z}_i)$
and $N/2$ anti-instantonic zero modes $\psi_{A}({\vec z}_j)$ the matrix
elements of the Dirac operator take the form (for zero quark mass)
\be
{\cal D}_{ILM} =
 \left (\begin{array}{cc} 0 & iT_{IA}\\
iT_{IA}^\dagger & 0 \end{array} \right )
\ee 
with the $N/2 \times N/2$ overlap matrix
given by
\be
 T_{IA} ~=~
 \langle\psi_{I}({\vec z}_i)|\,{\slash D}\,|\psi_{A}({\vec z}_j)\rangle
 ~\sim~ \rho_i \; \rho_j \, / \, |{\vec R}_{ij}|^3
\label{olpl}
\ee
for large separations ${\vec R}_{ij} = {\vec z}_i - {\vec z}_j$
between the center of instanton $i$ and anti-instanton $j$
with $\rho_i$ and $\rho_j$ their sizes.
Due to the chirality of the zero modes only matrix elements connecting 
 an instanton with an anti-instanton do not vanish. 
Physically the amplitude of the matrix elements $T_{IA}$ represents the 
 probability for a quark to hop between an instanton and an 
anti-instanton.
An isolated (anti-)instanton would cause an exact zero mode but
 this degeneracy is lifted through overlap with neighboring instantons.
Thus the would-be zero modes are effectively split around zero. 
As more (anti-)instantons are added a continuous band spectrum 
is formed with a spectral density finite at zero.
As mentioned previously, it is precisely 
this dynamically generated non-zero spectral density that causes,
 through the Bank-Casher relation, a finite value of the chiral condensate.    

\section{The chiral random banded model}

Here we study the spectral properties of an ensemble of
chiral random Hermitian $N\times N$ matrices given by
\be
\label{rb}
\mathcal{D}_{RBM} = \left (\begin{array}{cc} 0 & C\\
C^\dagger & 0 \end{array} \right )
\ee
where $C$ is a $N/2 \times N/2$ complex matrix with
independently distributed Gaussian variables with zero mean.
The variance of the matrix elements $C_{ij}$ are chosen to decay as a power of
$r = |i-j|$ which measures the distance from the diagonal.
Since the ILM uses periodic boundary conditions we use a periodic form
of the power-law decay \cite{evers} given by
\begin{equation}
\label{e4}
\langle |C_{ij}|^2\rangle =
\left\{ 1 + \left[
 \frac{\sin(2 \pi r/N)}{\pi b/N}
\right]^{2\alpha}\right\}^{-1}
\end{equation}
where $\alpha$ and $b$ are real parameters.
The choice of complex matrix elements corresponds to a matrix model with
a unitary symmetry which is appropriate for QCD with the phenomenologically
relevant $SU(3)$ color group.
Due to the chiral symmetry, the eigenvalues of (\ref{rb}) come in pairs of
$\pm \epsilon_i$.
This feature induces an additional level repulsion around zero 
 which results in different spectral correlations for eigenvalues near zero
 (the origin) and away from zero (the bulk).

In the bulk the spectral correlations should not be affected by the block 
 structure and should coincide with the non-chiral version of (\ref{rb}) which
  has been intensively studied in recent years \cite{prbm,evers}.  
  The use of the supersymmetry method \cite{efetov} permits an  
 analytical evaluation of both spectral properties and eigenfunction 
 statistics \cite{prbm} in a certain region of parameters.
In the thermodynamic limit the eigenfunctions
 are multifractal for $\alpha = 1$ and localized (delocalized)
 for $\alpha > 1$ ($\alpha < 1$) respectively \cite{prbm}. 
The spectral correlations in the $g=E_c/\Delta >> 1$
 ($E_c$ is the Thouless energy and $\Delta$ is the mean level spacing)
 limit can be expressed through the   
 spectral determinant of a classical diffusion operator \cite{andre}.
The two point correlation function is defined as
$R_{2}(s) = \Delta^2 \langle \rho(\epsilon)\rho(\epsilon + s\Delta) \rangle -1$
where $\rho(\epsilon)$ is the density of states at energy $\epsilon$
and the average is over an ensemble of RBM. 
For the unitary ensemble (our case)
\be
R_{2}(s) =
 - \frac{1}{4\pi^2}\frac{\partial^2}{\partial s^2} \ln \frac{D(s,g)}{s^2}
 + \frac{\cos(2\pi s)}{2 \pi^2 s^2} D(s,g) ~.
\label{kt}
\ee
Due to the power-law decay, the spectral determinant 
$D(s,g)= \prod_{n \neq 0}(1+s^{2}/\epsilon_{n}^2)^{-1}$ ( 
$\epsilon_{n}=g|n|^{2\alpha-1}$)  
corresponds with a process of anomalous
 diffusion \cite{prbm}. 
For $ 1/{2} < \alpha < 1$
the dimensionless conductance increases with the system size as
 $g = C_{\alpha}(b) N^{2-2\alpha}$ with $C_{\alpha}(b)$ a known constant.
The scaling 
 of $g$ thus resembles that of a   
 weakly disordered conductor in $d= {2}/{(2\alpha -1)}$  
 dimensions \cite{prbm}.  

\begin{figure}
\includegraphics[width=0.99\columnwidth]{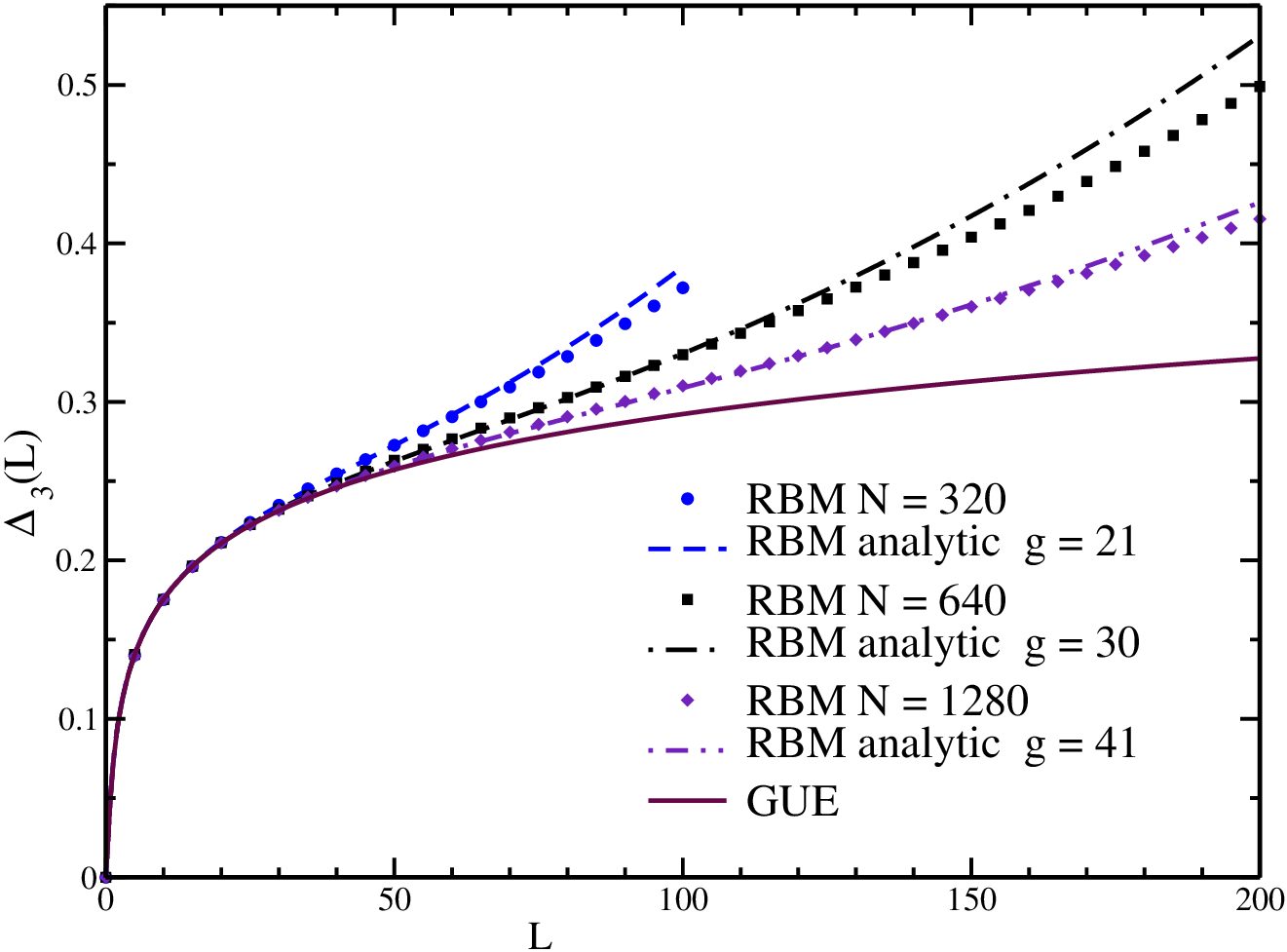}
\vspace{-6mm}
\caption{Spectral rigidity for the RBM in the bulk.
Points correspond to the numerical simulations of the chiral RBM
for $b = 1$.
Lines correspond to the analytical result of the
non-chiral RBM (\ref{kt}) for the given values of $g$.}
\label{d3bulkna} 
\vspace{-5mm}
\end{figure} 

The above results were derived for the non-chiral RBM.
As mentioned above they are also expected to describe the chiral RBM
 (\ref{rb}) in the bulk region.
We have chosen the scale of $b$ in (\ref{rb}) to agree with results for the
non-chiral model.
Strictly speaking, the factor $C_{\alpha}(b)$ was derived \cite{prbm}
 for $b \gg 1$ and it is not clear how the result should be modified for
 smaller $b$.
Below we will show that in fact the analytical results agree very well
even at $b=1$.
Unlike in the bulk, analytical results for the spectral correlations
close to the origin are not known.
They can be obtained by modifying the supersymmetry method to account 
for the chiral structure.
This issue will be postponed to a future publication \cite{kazu}
 and in this letter we will rely on numerical calculations for the study 
 of the spectral correlations close to the origin.
Lastly we remark that though the power-law decay in the instanton liquid
 is not in principle related to the 
 matrix index, as in our matrix model, 
 numerical simulations show that both models yield similar results \cite{par}.
We stick to (\ref{rb}) due to the availability  of analytical results.

Given the known results for the non-chiral RBM we can now choose the
parameter $\alpha$.
Recall that our motivation to propose a random banded model
is the decay of the ILM overlap matrix elements (\ref{olpl}).
It was shown in \cite{levitov1} that 
the spectral properties of systems with power-law hopping  
are similar in different dimensions provided that the decay 
exponent equaled the dimension.
Since in the ILM the decay exponent (three) is less than the dimension (four)
we expect this to map onto a 1D RBM model with $\alpha < 1$.
We choose $\alpha=3/4$ because
the volume dependence of the dimensionless conductance
(also called dimensionless Thouless energy)
in this case $g \approx 1.17\sqrt{b N}$ \cite{prbm}
coincides with what is expected for QCD
according to chiral perturbation theory \cite{gasser}.
As mentioned above, the spectral properties
of the RBM at $\alpha=3/4$ are similar to those of a disordered 
conductor in four dimensions.

\begin{figure}
\includegraphics[width=0.99\columnwidth]{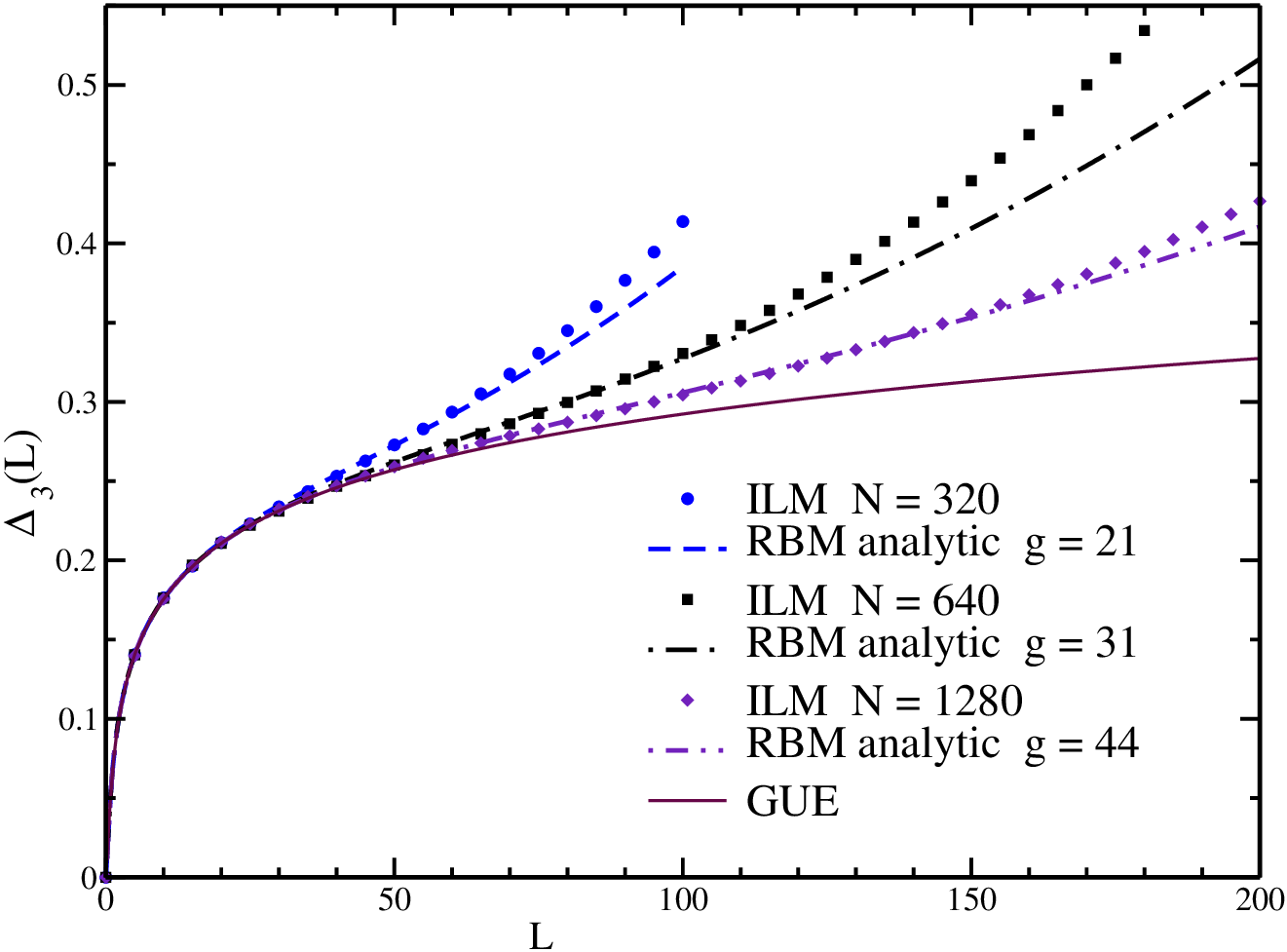}
\vspace{-6mm}
\caption{Spectral rigidity for the ILM and RBM in the bulk.
Points correspond to numerical simulation of the ILM.
Lines correspond to the analytical result of the non-chiral RBM (\ref{kt})
for the given values of $g$.}
\label{d3bulkia}
\vspace{-5mm}
\end{figure} 

\section{Results}
\vspace{-4mm}

We generated sets of matrices for the RBM and the ILM and calculated their
eigenvalues in order to compare the spectral correlations.
The results of each simulation are averaged over $10^4$ configurations.
For both models the simulations were done in the quenched approximation
which allowed us to use values of $N$ ranging from 320 up to 1280.
For the ILM we used the same model studied in \cite{VO} including the
standard density of $N/V=1 \fm^{-4}$.
Additional details can be found in \cite{SS97}.

We first check that the result (\ref{kt}), derived for the non-chiral
RBM, agrees with the chiral result in the bulk.
The analytical formula is actually divergent for $\alpha=3/4$ and needs
a cutoff in the spectral determinant $D(s,g)$.
If we calculate the number variance which is given by
$\Sigma^{2}(L)=L +2\int_{0}^{L}(L-s)R_{2}(s)ds$
we find that it is strongly cutoff dependent.
Instead we look at the spectral rigidity
$\Delta_{3}(L)=\frac{2}{L^4}\int_{0}^{L}(L^3-2L^2x+x^3)\Sigma^{2}(x)dx$
which is not sensitive to the cutoff.

In Figure \ref{d3bulkna} we show the spectral rigidity in the bulk 
of the chiral RBM obtained from numerical simulation at $b=1$ along with
the analytical formula of the non-chiral model for different values of $g$.
The values of $g$ were chosen by eye to match the numerical
results for the corresponding $N$.
We find good agreement with the theoretical expectation
$g \approx 1.17 \sqrt{N}$ within about $2\%$.

In Figure \ref{d3bulkia} we compare the spectral rigidity in the bulk
of the ILM with analytical predictions of the RBM.
Again the values of $g$ for each volume were chosen by eye to closely match
the ILM.
The analytical results using these values of $g$ agree
well with the ILM up to a scale of about $30\%$ of the number of positive
eigenvalues ($N/2$).
The values of $g$ scale closely to $g \approx 1.23 \sqrt{N}$ 
except for the smallest volume ($N=320$) where a $5\%$ deviation is observed.
This corresponds to $b \approx 1.1$.
The observed scaling of the dimensionless Thouless energy 
in the chiral RBM is in agreement with theoretical predictions
\cite{gasser,VO} and lattice results \cite{tilo} for QCD.

\begin{figure}
\includegraphics[width=0.99\columnwidth,clip]{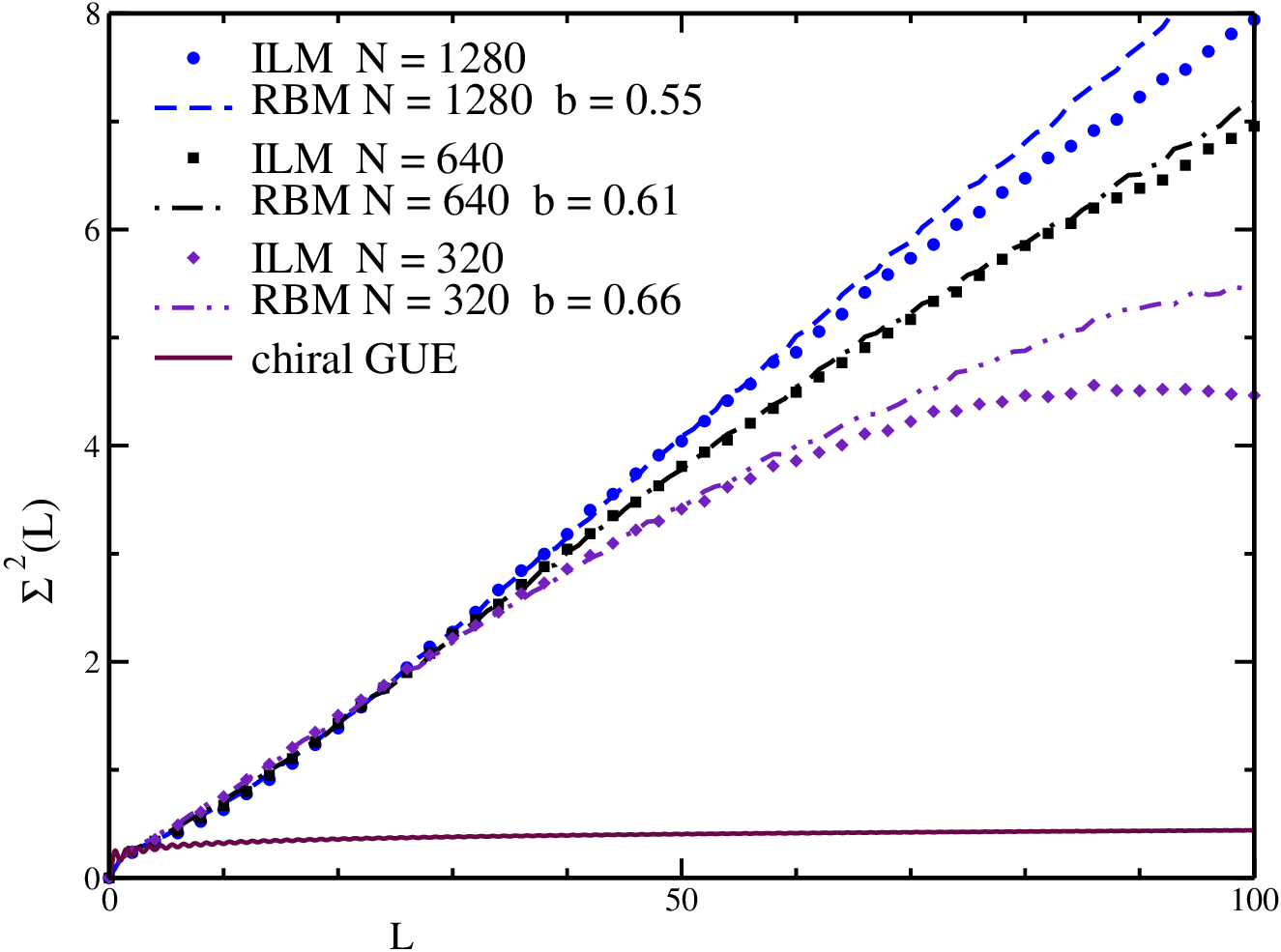}
\vspace{-7mm}
\caption{Number variance $\Sigma^2(L)$ close to the origin.
Points correspond to numerical simulations of the ILM.
Lines correspond to numerical simulations of the chiral RBM
for the given values of $b$.}
\label{nvorigin}
\vspace{-7mm}
\end{figure} 

We do not yet have analytical results for the spectral correlations 
of the RBM at the origin and therefore rely solely
on numerical simulations. 
In Figure \ref{nvorigin} we show the number variance near the origin
for the ILM and chiral RBM.
The corresponding values of $b$ for the RBM were also chosen by eye to
provide a good fit.
The agreement between both models is very good up to about 50 eigenvalues
 but we could not fit all volumes with a single $b$. 
The reason is that the chiral RBM 
 has the same $g \sim \sqrt{N}$ scaling as in the bulk
 while the instanton liquid shows a weaker volume dependence.
We find the scaling $b \approx 1.4 \, N^{-0.13}$
 which gives $g \approx 1.4 \, N^{0.43}$ for the ILM.
It would be interesting to compare these results with lattice simulations
where a $g \sim \sqrt{N}$ scaling has also been reported close to the origin.

Finally we mention how the chiral RBM 
should be modified to describe QCD at finite temperature. 
As usual in field theory, temperature is introduced
 by compactifying one of the spatial dimensions. 
Thus the effect of temperature in Euclidean QCD is to 
 reduce the effective dimensionality 
of the system to three. Now since 
the effective dimension of the space
 matches the power-law decay of the QCD Dirac operator ($\sim 1/R^3$)
  one expects, according to \cite{par,levitov1}, 
 multifractal wavefunctions typical of a metal-insulator transition.
 The same chiral RBM proposed in this paper may be used 
 in this situation but with $\alpha=1$ (see \cite{ant1} for a model 
 with similar spectral properties).
The above arguments suggest that if the restoration of chiral symmetry  
at finite temperature is dominated by instantons, the 
physical mechanism leading 
to the quark-gluon plasma state of matter would be      
similar to a metal-insulator transition.
Clearly further work is needed to explore this exciting relation.

To conclude, we have proposed a 
chiral random banded model with power-law decay in order to 
describe the spectral correlations of the QCD Dirac operator
beyond the Thouless energy.
We have thus combined the asymptotic power-law tail 
observed in instanton liquid models 
with the random matrix approach valid for small spacings.
We have provided numerical evidence that the resulting chiral RBM  
does (at least for the two-point function)
describe the spectral correlations of the
QCD Dirac operator well beyond the Thouless energy. 
Finally we have mentioned that at the 
finite temperature chiral restoration transition
the appropriately modified chiral RBM predicts a metal-insulator behavior
including multifractal wavefunctions
and the physics of the Anderson transition in the QCD vacuum.              

We thank J.J.M~Verbaarschot for illuminating discussions. A.M.G. was
 supported by the EU network ``Mathematical aspects of quantum chaos''.
J.C.O. was supported in part by NSF PHY 01-39929.

\end{document}